# Complex interplay between 3d and 4f magnetic systems and magnetic chirality in multiferroic $Dy_{1-x}Ho_xMnO_3$ (x = 0, 0.2)

A.N. Matveeva[1], I. A. Zobkalo[1], A. Sazonov[2,*], M. Meven[2], A.L. Freidman[3], S.V. Semenov[3], M.I. Kolkov[3], K. Yu. K. Terentjev[3], N. S. Pavlovskiy[3], K. A. Shaykhutdinov[3], V. Hutanu[2,#]

[1] *Petersburg Nuclear Physics Institute by B.P. Konstantinov of NRC «Kurchatov Institute», 188300 Gatchina, Russia.*

[2] *Institute of Crystallography, RWTH Aachen University and Jülich Centre for Neutron Science at Heinz Maier- Leibnitz Zentrum, Garching, Germany*

[3] *Kirensky Institute of Physics, Federal Research Center, Krasnoyarsk 660036, Russia*

[*] *Current affiliation: ESS Data Management & Software Centre (DMSC), 2200 Copenhagen, Denmark*

[#] *Current affiliation: Technical University of Munich, ZWE FRM II, 85748 Garching, Germany*

Structural, magnetic and multiferroic properties of single crystals of $Dy_{1-x}Ho_xMnO_3$ (x = 0, 0.2) were investigated by the different methods of polarized and classical neutron diffraction and macroscopic methods in order to determine the effect of Ho doping on the magneto-electric behavior of the title compounds. It is shown that substitution by Ho of 20% on the position of Dy do not change overall crystal symmetry of compound. It remains of *Pnma* type for both compositions down to the very low temperatures. Magnetic ordering don't change the crystal structure. Precise magnetic order and it detailed temperature and field evolution both in the pristine and substituted compounds we determined using single crystal neutron diffraction and magnetization measurements. The results show a complex interplay between transition metal and rear earth magnetic sub lattices leading to so-called "Mn-controlled" and "Dy- controlled" magnetic states. Using polarized neutron diffraction 3D character of rear earth magnetic order in $Dy_{0.8}Ho_{0.2}MnO_3$ in contrast to $DyMnO_3$ and occurrence of the chiral type magnetic structure on Mn subsystem could be revealed. The influence of the external electric field on the magnetic chirality could be directly evidenced, proving strong magneto-electric coupling in multiferroic phase. The study of the electric polarization under similar temperatures and fields on the same samples provides the direct correlation between the results of the microscopic and macroscopic investigations.

I. INTRODUCTION.

Compounds from the orthorhombic manganites family $RMnO_3$ (where R = Tb, Dy, Gd, Ho, Y) are considered to be a typical representatives of the multiferroics of the type II [1–6]. Magnetic ordering in these compounds reduces the symmetry of a nonpolar parent phase to a polar magnetic one, and magnetostructural coupling leads to the appearance of an electrically polar state, thus inducing improper ferroelectricity. Close coupling between the magnetic and ferroelectric orders in multiferroics of type II offer the possibility to manipulate one ferroic property by the field of the other one (e.g. electric polarization by magnetic field or magnetic chirality by electric field), thus opening the ways for the development of the new solid state devices. The understanding of microscopic mechanisms of the improper magnetoelectricity in multiferroic materials in general and in $RMnO_3$ series in particularly attracted much scientific interest to these materials in the last years. Of great interest is DyMnO$_3$ (DMO), which develops one of the largest electric polarization among these type of manganites. 3D magnetic order in DMO is established at $T_N$ = 39 K by the magnetic ordering of Mn$^{3+}$ ions in spin wave structure. An additional transition from non-chiral longitudinal spin density wave modulation to the chiral cycloidal-type structure takes place at $T_{Ch}$ = 19 K[3]. Rare earth (RE) magnetic system becomes polarized at comparatively high temperature by the exchange field of the Mn subsystem and have the same propagation vector, as the Mn one: $\mathbf{k}$ = (0 $k_y$ 0). Remarkably the emergence of ferroelectric order coincides with the transition of the Mn magnetic order from the non-chiral type to the chiral one. The relationship between ferroelectric polarization and the cycloidal magnetic structure can be satisfactorily described in the framework of the inverse Dzyaloshinsky-Moria (DMI) model [7, 8]. It is worth noting here that Dy$^{3+}$ spontaneous ordering into a commensurate antiferromagnetic structure $\mathbf{k}_{Dy}$ = (0 1/2 0), which takes place at ~ 7 K, reduces the ferroelectric polarization significantly [3].

A similar situation concerning the evolution of the magnetic ordering takes place also in HoMnO$_3$ and YMnO$_3$ manganites with the only difference that the transverse spin wave structure changes to an antiferromagnetic structure of E-type i.e. an up-up-down-down order at about 27 K [9]. For these compounds a mechanism of ferroelectric polarization, based on the exchange-striction is proposed. It is predicted that this mechanism generate a much larger ferroelectric polarization than the previous one [4-6].

Despite the large number of works on this topic, the precise microscopic mechanism for the occurrence of spontaneous electric polarization in these systems is still under debate. The models of inverse DMI were developed based on the early experimental findings on TbMnO$_3$ [7, 8, 13]. In these models, the polarization is proportional to the vector product of the adjacent Mn spins solely $\mathbf{P}_e$ ~ [$\mathbf{S}_{iMn} \times \mathbf{S}_{jMn}$], and it describes the multiferroic phenomena in RMnO$_3$ rather well. The later subsequent studies on the TbMnO$_3$ and especially on DyMnO$_3$ suggest however, that an additional mechanism of polarization emergence may be involved in these compounds [14-17]. Nowadays it is generally assumed, that symmetric exchange-striction between neighboring Dy and Mn ions lead to the enhancement of ferroelectric polarization in DyMnO$_3$: $\mathbf{P}_e$ ~ ($\mathbf{S}_{Dy} \cdot \mathbf{S}_{Mn}$). Thus, the total polarization along the $c$-axis emerges due to two different microscopic mechanisms: from the symmetric exchange-striction between the Dy and Mn ions and from the action of inverse DMI between the Mn ions sublattice. An important issue in this regard is the coherence of the Mn$^{3+}$ and R$^{3+}$ magnetic phases, which plays a crucial role in generating the polarization of the symmetric exchange-striction origin [15-17]. The non-collinear Mn moment alignment results in a ferroelectric state, and then the exchange-striction mechanism enhances the resulting polarization. The chiral incommensurate magnetic ordering of manganese sublattice in DyMnO$_3$ can be regarded as essential prerequisite for the polarization enhancement by the exchange-striction, similar to the situation proposed in the RMn$_2$O$_5$ manganites [18]. It is worth mentioning, that after formation of the commensurate magnetic (CM) order of the Dy sublattice below $T_N^{Dy}$ (Dy-Dy interaction becoming dominant), electric polarization is strongly suppressed, underlaying the importance of the Mn-Dy interactions as the driving mechanism in the polarization enhancement.

In HoMnO$_3$ the strong Ho$^{3+}$- Mn$^{3+}$ coupling is reported to occur at higher temperatures resulting in the fact that both Mn and Ho sublattices have same CM order below ~27 K and preserves the coherence between the Ho$^{3+}$ and Mn$^{3+}$ magnetic systems down to the very low temperatures [9]. Thus, the polarization arising from the Ho$^{3+}$-Mn$^{3+}$ exchange-striction exists down to lowest temperatures of the same magnitude as that of Dy$^{3+}$-Mn$^{3+}$ above the $T_N^{Dy}$ in DMO. Therefore, the sequential substitution of Dy$^{3+}$ by Ho$^{3+}$ allows one to control the entire spin structure (not just the spin structure of the RE subsystem) and to adjust the multiferroicity of RMnO$_3$ by changing the RE ions [21-25]. The recent studies of substituted compounds Dy$_{1-x}$Ho$_x$MnO$_3$ (DHMO) demonstrate the enhancement of ferroelectric polarization at low temperatures, which implies the preservation of the coherence between the R$^{3+}$ magnetic subsystem with the Mn$^{3+}$ spins in the cycloidal magnetic phase [23] below the $T_{NR}$. The complete suppression of the Dy-ordering-related polarization drop-down could be reached by the doping level of $x = 0.2$. Moreover, the Ho-substituted powder samples showed higher polarization in comparison to the parent pure DMO. Such a substitution seems to be a promising technological way of both the enhancement of the polarization magnitude itself, but also to extend the temperature region where the stable polarization could be exploit. The details about the magnetic order in the mixed compound DHMO remains however unstudied. Moreover, the direct correlation between the observed enhancement of the electric polarization and the preservation of the cycloidal order is still missing. In order to study the coupling between the RE and Mn magnetic subsystems in Dy$_{1-x}$Ho$_x$MnO$_3$ ($x = 0, 0.2$) multiferroics in detail and to clear the effect of the RE ion type on the magnetic order and subsequent electric polarization, an exhaustive comparative single crystal neutron diffraction experiments (including polarized neutrons diffraction) were carried out in this compounds. The macroscopic measurement of electric polarization in the same samples are important point of this study in order to establish a direct link between microscopic details on the magnetic structure to the macroscopic properties.

It is important to note that RMnO$_3$ compounds, depending on the ionic radius of the RE, have either orthorhombic perovskite structure or hexagonal one under normal growth conditions. DyMnO$_3$ is just located at the boundary of two crystal modifications and, depending on the growth conditions crystalizes in one of these two structures [19, 20]. Thus, the RE not only affects the multiferroism of RMnO$_3$, but could also directly influence the crystal symmetry. As the known crystal symmetry is an important prerequisite for the determining of the magnetic structure, detailed crystallographic study of the both pure and substituted compounds has been performed by the means of the single crystal neutron diffraction.

## II. EXPERIMENTAL.

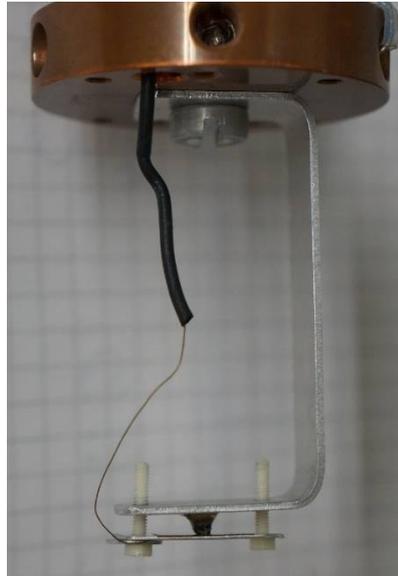

FIG. 1. Dy$_{0.8}$Ho$_{0.2}$MnO$_3$ single crystal placed between two aluminum electrodes on the cold finger of cryostat.

High quality single crystals of Dy$_{1-x}$Ho$_x$MnO$_3$ (x=0, 0.2) were grown by spontaneous crystallization from a solution in a melt. The ratio of the chemical reagents PbO:PbF$_2$:B$_2$O$_3$ used as a solvent was 0.84:0.14:0.01 respectively. The ratio of the desired composition $Dy_{1-x}Ho_xMnO_3$ to the solvent was 1:9. High purity reagents 99.9% (Alfa Aesar) of $PbO$, $PbF_2$, $MnO_2$, $Dy_2O_3$, $Ho_2O_3$, $B_2O_3$ were used for synthesis. The technological regime consist of heating the charge in a platinum crucible to 1200°C, followed by cooling for two weeks to a temperature of 700°C. The obtained single crystals have the shape of parallelepipeds with sizes up to 1.5×1.5×3 mm$^3$, shiny surfaces and black color. The synthesis and characterization of the orthorhombic single crystals of the family $Dy_{1-x}Ho_xMnO_3$ (x=0 - 0.4) were described in detail in our previous work [26].

For the electric polarization measurements plate-like sample in the form of a flat capacitor was prepared. The capacitor plates were formed from an epoxy-based conductive paste with a silver filler deposited onto the preliminary polished crystal faces perpendicular to the crystallographic *a*-axis. The electric polarization was determined by measuring the electric charge flowing-off the capacitor plates with a Keithley 6517B electrometer.

Measurements of the magnetic properties were carried out at the Kirensky Institute of Physics on an in-house-made vibration magnetometer with a superconducting solenoid. This magnetometer is designed for precision measurements of magnetic characteristics of various materials using an automated measuring system. The device allows to measure the magnetization as function of temperature M(T) in the range of 4.2-300 K and an applied magnetic fields up to 6 T M(H).

The neutron diffraction studies were performed at diffractometers HEIDI [27] and POLI [28] at the Heinz Maier-Leibnitz Zentrum (MLZ, Garching). The four-circle diffractometer HEIDI is designed for detailed studies of the structural and magnetic properties of single crystals with hot neutrons. While the single crystal diffractometer POLI is specialized on different methods of polarized neutron diffraction. In order to obtain quantitative chiral characteristics of magnetic structures the method of spherical neutron polarimetry (SNP) [29] was used. It allows to determine all components of the scattering polarization matrix. SNP technique is implemented on POLI using third generation zero-field Cryogenic Polarization Analysis Device CRYOPAD [30]. The unique feature of POLI is that it uses $^3$He spin filters for neutron polarization and analysis. Such a setup is very efficient for the short wavelength neutrons in order to increase the angular resolution and the flux density of the polarized neutrons on the sample position. On the other side, the time dependence of the polarizing (analyzing) efficiency of the filters represents a main drawback of the technique, since the precise knowledge of the incoming and scattered beam polarization is essential for the SNP. This requires some additional corrections, which may lead to a decrease in the statistical accuracy of the measurement. The standard correction procedure described elsewhere [31] was applied to the SNP data.

Measurements of the chiral scattering in the external electric field were made in a "half polarized" mode, i.e. without analysis of the polarization for the scattered neutrons. This mode is less time consuming than SNP because it avoids the transmission of the analyzer, but still gives the possibility to measure the pure chiral scattering by *xx* components of the polarization matrix. The results of these measurements were scaled then with those from SNP data. In order to apply an electric field during measurement, the crystal was glued between two aluminum plates forming kind of electric capacitor using non-magnetic, electrically isolating ceramic screws [Fig.1]. The electric field was applied along the *c*-axis in the crystal. The sample was cooled down using dedicated long-tale top-loading closed-cycle cryostat model Janis SHI-950-T. The essentials of the measurements with high electric fields in the cryogenic environments consist in fine-tuning the pressure of helium exchange gas continuously along a thin line where it is possible to apply several kV/mm voltage on the sample without an electric breakdown maintaining the necessary temperature of the sample at the same time. A reliable setup for the in-situ pressure control and regulation within a cryostat has been developed and calibrated before the experiment was started.

In all SNP experiments we employed the coordinate system which is conventional for the polarized neutron in which the instrumental *x*-direction coincides with the scattering vector *q*, the *y*-direction lies in the scattering plane and is perpendicular to *x*, and the *z*-axis is directed vertically forming the right-handed Cartesian system. The crystal orientation was chosen as follows: *c*-axis was oriented vertically, *a* and *b* axes lying in a horizontal scattering plane.

## III. RESULTS AND DISCUSSIONS

### A. Crystal structure of $Dy_{1-x}Ho_xMnO_3$ (x = 0, 0.2).

Generally, in the literature the perovskite orthorhombic space group *Pbnm* is assumed for the crystal structure of DMO both at room and low temperatures. However, recent detailed study by high resolution neutron and synchrotron powder diffraction at low temperatures, suggested a lowering of the crystal symmetry along with magnetic transitions to the ICM and CM phases to space groups *Pna2₁* or *P2₁* [33]. In order to check for these suggested fine atomic displacements, related to the occurrence of the electric polarization in DMO, we performed a number of single crystal neutron diffraction data collects at various low temperatures: 50 K paramagnetic, 25 K Mn-ICM, 8 K – ordering of Dy, and 2.4 K both Dy and Mn ordered and lowest available within used cryostat. The crystal structure of DMO was explored at the single crystal diffractometer HEIDI with wavelength λ = 0.556 Å in order to reduce the parasitic effect of absorption on Dy. The crystal structure refinement then was performed using FullProf suite [32]. According to our results, the orthorhombic space group *Pbnm* [Fig. 2] describes well the crystal structure in DMO at all these temperatures. Neither significant refinement quality or goodness of fit changes in dependence on the temperature, no crystal structure parameters change was observed. Both fractional atomic positions for all atoms and isotropic/anisotropic displacement parameters differ by less than one standard deviation between paramagnetic and highly ordered magnetic temperatures. The obtained atomic positions for 2.4 K are shown (in Table I). The presented here crystal structure parameters obtained at the low-temperature single crystal neutron diffraction are in a good agreement with those reported for DMO previously by other methods like neutron powder diffraction [33] or single crystal X-ray diffraction [34]. We did not detect neither additional Bragg reflections, forbidden in *Pbnm* group, nor splitting of the measured reflections. We consider that most probably, because of the limitation of the method (lower resolution using hot neutrons), our results could not either confirm, no disaffirm the newly suggested in [33] fine atomic displacement and related symmetry lowering. On the other side, our results clearly show, that traditionally considered space group *Pbnm* still describes well the average symmetry in DMO at low temperatures in different magnetic states.

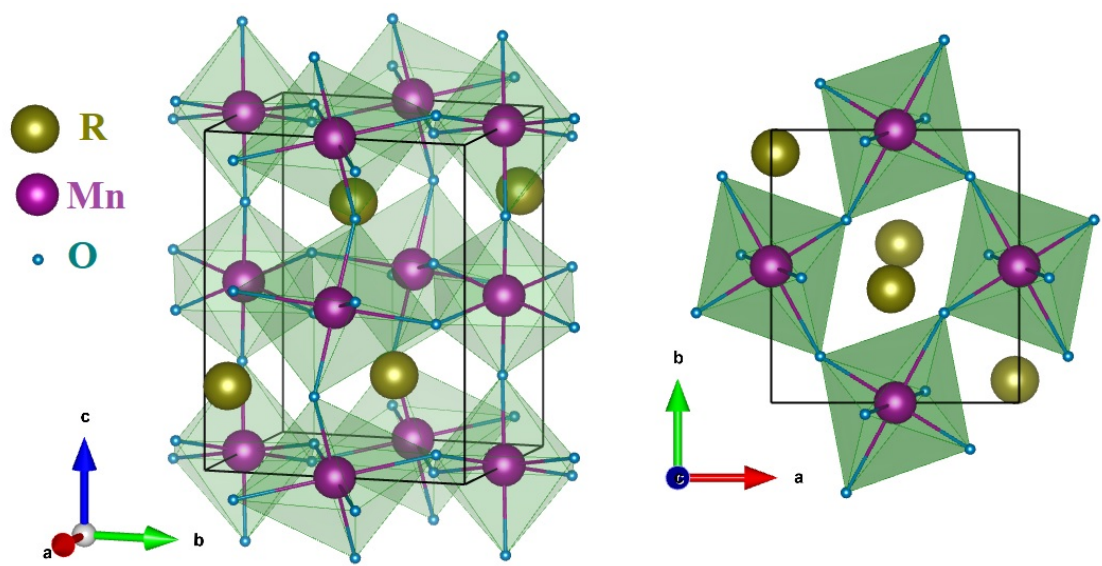

FIG. 2. Typical RMnO$_3$ distorted perovskite crystal structure within space group Pbnm.

For the crystal structure refinement of DHMO 542 nuclear Bragg reflections were measured on the single crystal diffractometer POLI at wavelength λ = 0.897 Å in non-polarized mode. Two data collections were performed: at room temperature and at 4 K respectively. Among measured data 179 independent Bragg reflections satisfying the condition I >3σ(I) were used for the structural refinement. Obtained for substituted compound DHMO results reveal that space group *Pbnm* describes well it crystal structure at both temperatures. Same as in DHO on HEiDi also in DHMO, only reflections allowed in *Pbnm* space group were detected and no splitting within the resolution of POLI diffractometer observed. Atomic positions in DHMO resulting from the refinement at 300 and 4 K respectively are shown in (Table I). Like it was in a previous case for DMO, no significant difference was detected in the unit cell parameters as well as in the atoms coordinates for low temperature, below $T_N$. As it can be seen from (Table I) that the structural parameters for the both "pure" DMO and substituted DHMO are very similar. Thus, the substitution by Ho of the Dy at level of 0.2 do not change the overall crystal structure. Assuming that Ho occupies the same as Dy Wyckoff position 4c and the total occupation of the RE site is one, the degree of the replacement of Dy by Ho was checked when refining the crystal structure of DHMO. The resulting from the structural refinement occupancies for Dy of 0.79(3) and Ho 0.21(3) respectively for the 4c site, could be obtained in the perfect agreement with the expected molar ratio from the crystal growing.

TABLE I. Fractional atomic positions in DMO and DHMO at various temperatures resulted from the structural refinement in space group Pbnm, determined by single crystal neutron diffraction..

| | DyMnO$_3$, 2.4 K a = 5.269(7), b = 5.845(4), c = 7.331(4) Å | | | Dy$_{0.8}$Ho$_{0.2}$MnO$_3$, 300 K a = 5.27(1), b = 5.840(4), c = 7.360(8) Å | | | Dy$_{0.8}$Ho$_{0.2}$MnO$_3$, 4 K a = 5.268(3), b = 5.835(4), c = 7.358(1) Å | | |
|---|---|---|---|---|---|---|---|---|---|
| Atom | x | y | z | x | y | z | x | y | z |
| Dy/Ho, 4c | -0.0180(2) | 0.0829(2) | 0.25 | -0.0169(5) | 0.0820(1) | 0.25 | -0.0178(4) | 0.0830(9) | 0.25 |
| Mn, 4b | 0.5 | 0 | 0 | 0.5 | 0 | 0 | 0.5 | 0 | 0 |
| O1, 4c | 0.1088(9) | 0.4622(4) | 0.25 | 0.1083(3) | 0.4640(4) | 0.25 | 0.1093(3) | 0.4622(2) | 0.25 |
| O2, 8d | 0.7022(5) | 0.3276(5) | 0.0527(6) | 0.7020(4) | 0.3285(5) | 0.0513(6) | 0.7013(9) | 0.3272(4) | 0.0525(5) |
| | $R_F$ = 4.49 | | | $R_F$ = 1.94 | | | $R_F$ = 1.51 | | |

### B. Temperature evolution of the magnetic structure.

Investigations of the magnetic structure of DMO were performed at neutron diffractometer HEIDI. Measurements were carried both in cooling and heating mode. Thermal evolution of the magnetic structure of DMO can be conveniently traced on the thermal dependence of the satellite (0 $k_y$ 1). The onset of the magnetic ordering was observed at $T_N^{Mn} \approx 38$ K with propagation vector $\boldsymbol{k}^{Mn}$ = (0 $k_y$ 0), $k_y$ = 0.355(2), which is in a good agreement with previous works [3, 14]. While cooling below $T_N^{Mn}$, the intensity of this satellite increases gradually down to ~ 19 K, and then a steep increase of the intensity is observed by further temperature decrease [Fig. 3(a)]. This rapid increase is connected to the ordering of the Dy magnetic sublattice induced by Mn ordered sublattice with the same propagation vector $\boldsymbol{k}^{Mn}$ as observed from the resonant X-ray magnetic scattering (XRMS) on the L-edge of Dy [14, 35, 36]. Our observations from [Fig. 3(a)] are in a good agreement with previously reported synchrotron data, but also neutron scattering results [37]. Remarkably, this (19 K) temperature coincides with the transition from longitudinal spin wave order of Mn to the cycloidal one and also to the temperature $T_{CE}$ = 19 K where the ferroelectric transition occurs [2, 3]. Below $T \sim 6.5$ K the intensity of the ICM satellite drops sharply to the value similar to that observed at temperatures above 27 K characteristic to the ordered Mn lattice only. At this temperature $T_N^{Dy} \sim 6.5$ K the spontaneous ordering of the rare earth magnetic system into an independent CM AFM structure with propagation vector $\boldsymbol{k}^{Dy}$ = (0 0.5 0) takes place. In cooling mode observed in our crystal ICM wave vector component $k_y$ related to the Mn ordering keeps its initial value $k_y$ = 0.355(2) unchanged between 38 and ~ 28 K in good agreement to the previous neutron diffraction study [37]. This behavior is somehow different to that resulting from the observation of the crystal lattice modulation peak with double vector component 2*$k_y$ by synchrotron scattering [14, 35, 36]. In the mentioned synchrotron studies a continuous change of $k_y$ between the $T_N^{Mn}$ and $T_{CE}$ is reported; this difference can be traced to the fact that those measurements were performed in the heating mode, which we will discuss later. Below 28 K $k_y$ starts gradually change from 0.355 to 0.37 at 19 K.

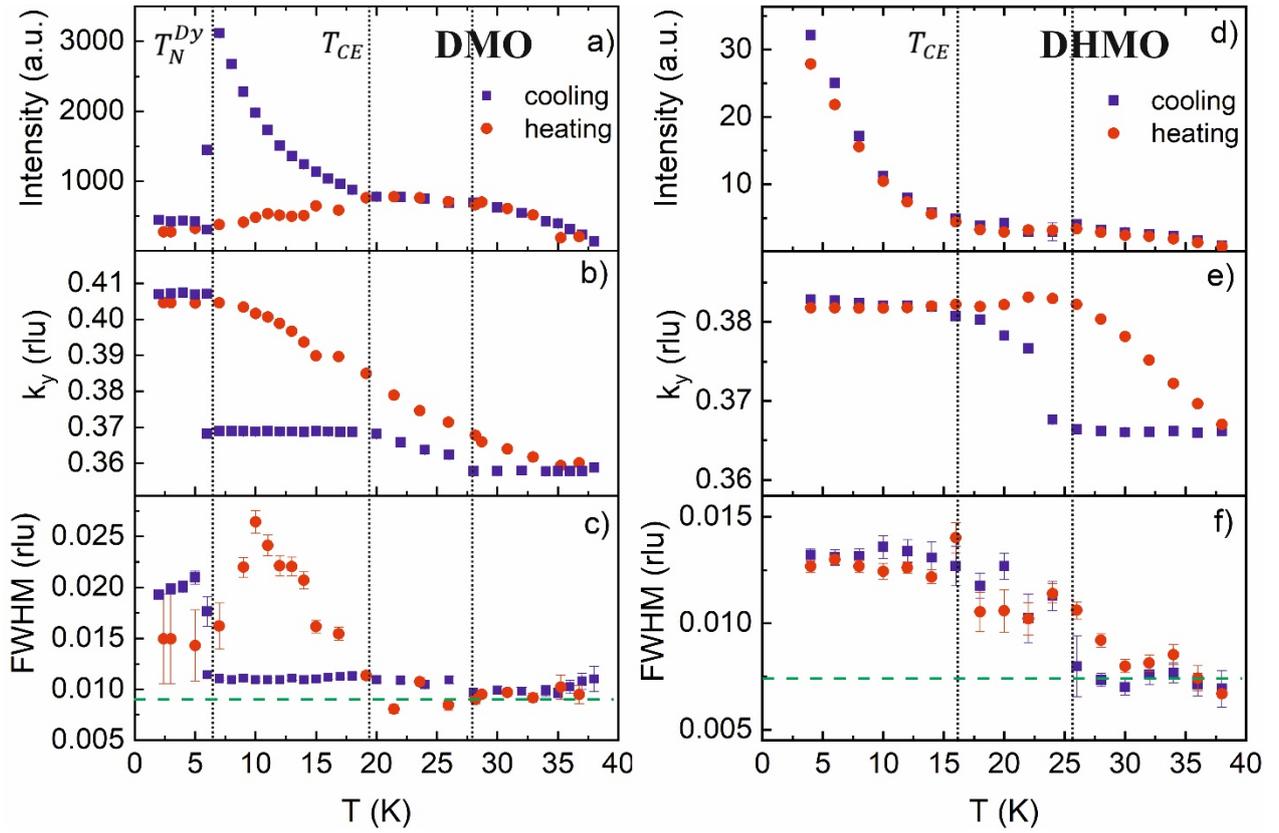

FIG. 3. Temperature evolution of the ICM magnetic Bragg reflection (2 $k_y$ 1) in DyMnO$_3$ and Dy$_{0.8}$Ho$_{0.2}$MnO$_3$ samples respectively by cooling and heating. a) and d) panels show the integrated intensity of the peaks; b) and e) the wave vectors evolution; c) and f) panels: full widths at half maximum (FWHM) of the magnetic peaks. The horizontal lines in the panels c) and f) denote the resolution limit of the instrument by the measurement. The vertical broken lines indicate the observed characteristic magnetic phase transition temperatures.

Below this lock-in temperature, the wave vector stay constant while cooling down to $T_N^{Dy}$, where it jumps abruptly to value $k_y = 0.405(2)$ [Fig. 3(b)]. The constant $k_y$ value between $T_{CE} \approx 19$ K and $T_N^{Dy} \approx 6.5$ K agrees well with previous synchrotron and neutron results. However, to the best of our knowledge, the sudden change of the $k_y$ to the value 0.405 below $T_N^{Dy}$ hasn't been reported previously. In the detailed neutron diffraction study of the Mn sublattice in the DMO [37] a splitting of the $k_y$ into two peaks with $k_{y1} = 0.363$ and $k_{y2} = 0.39$ was observed. Interestingly, that crystal lattice modulations with ICM wave vector $\boldsymbol{k}_l = \boldsymbol{k}^{Dy} + \boldsymbol{k}^{Mn}$, which is induced by a magnetoelastic coupling between Dy and Mn magnetic moments was found below $T_N^{Dy}$ [35]. Observed on single crystals by synchrotron scattering lattice distortion value $k_l = 0.905$ is in a perfect agreement with reported here magnetic vector components $\boldsymbol{k}^{Mn} = 0.405$ and $\boldsymbol{k}^{Dy} = 0.5$ determined below $T_N^{Dy}$ from the neutron diffraction. The observed (in Ref. 37) deviation from that value was attributed to the use of the large mosaic single crystal sample with 90% isotopically substituted $^{162}$Dy. In such a crystal some parts of the sample may still remain not perfectly ordered down to 4 K. Our observation is a direct evidence not only for the strong magnetoelastic coupling of the both Mn and Dy magnetic sublattices with the crystal lattice, but also for the existence of an active interaction between the two magnetic sublattices, which should be considered for the description of the multiferoicity mechanism in DMO.

The thermal dependence of the magnetic satellite width is displayed [in Fig. 3(c)]. While cooling, peaks full width at the half maximum (FWHM) stay constant down to $T_N^{Dy} \sim 6.5$ K, when it sharply increases to almost double resolution-limited value. The broadening of the ICM peaks in DMO below $T_N^{Dy}$ was observed also by the XRMS experiments and was attributed to presence of the narrow dispersion of the wave vectors in the sample rather than the short range correlations [14]. Taking into account that FWHM does not change by further temperature reduction down to 2.5 K our results obtained by neutron diffraction support this assumption. Worth mentioning is, that our limited resolution by using very short neutron wavelength diffraction could not permit the observation of the ICM peak splitting.

In heating mode the temperature dependence of the integrated intensity demonstrates a noteworthy hysteretic behavior comparing to the cooling case. Looking at the intensity dependence [Fig. 3(a)], one can see that it increases only very slowly above $T_N^{Dy}$ and returns to its previous "cooling mode" value only above ~ 19 K (remarkably coinciding with $T_{CE}$). A qualitatively similar (but quantitatively much less pronounced) hysteretic behavior in the intensity of the magnetic satellite (0, 1, 2)+ around 7 K related to the occurrence of the Dy magnetic ordering was reported also from XRMS measurements [14] in principal agreement with our results. The observed hysteresis resembles well the results presented (in Ref. 37) obtained also by the single crystal neutron diffraction. This hysteretic behavior is an additional manifestation of a strong magnetic coupling between Mn and Dy. In cooling mode the Mn subsystem polarizes the Dy down to $T_N^{Dy}$, and polarized Dy subsystem has coherent propagation with Mn one, thus this state we will call the "Mn-controlled" one. Then, below $T_N^{Dy}$, the RE subsystem already has a significant impact on the Mn magnetic ordering, the so cold "Dy-controlled" state established. In the cooling mode this can be shown by the shift of the incommensurate propagation vector and by the broadening of the incommensurate satellite peaks below $T_N^{Dy}$. In the heating mode the Dy magnetic subsystem preserves its impact on the Mn even after loosing its own CM long range order. It disturbs the Mn magnetic subsystem up to ~ 19 K, which leads to a broadening of the satellites in this region [Fig. 3(c)]. In addition, the Mn propagation vector $\boldsymbol{k}_y^{Mn}$ has larger values than that during cooling mode almost up to $T_N^{Mn}$ [Fig. 3(b)]. It seems that the "Dy-controlled" state while heating lasts up to higher temperatures than "Mn-controlled" one while cooling. Figure 4 shows the temperature dependence of the CM AFM reflections, originating from the $Dy^{3+}$ spontaneous ordering below $T_N^{Dy}$. The integrated intensity of (0, 0.5, 2) reflection shows a regular hysteretic loop with a width of about 1.8 K between cooling and heating branches in good agreement with previously reported 1.7 K [37]. Taking as transition temperature the inflection point, results in the $T_N^{Dy}$-cooling ≈ 6.1 K and $T_N^{Dy}$-heating ≈ 7.9 K. These values are in the good agreement with the most of the previously reported $T_N^{Dy}$ as well. In some published works, an additional critical temperature of about 12 K attributed to polarization flop transition [38] in DMO is reported. It is worth mentioning, that our single crystal neuron diffraction data on CM and ICM peaks from Mn and Dy magnetic ordering do not show any significant anomalies at that temperature. The only possible trace of such transition in our data may be the broadening of the ICM peak at about 10-12 K while heating [red symbols in the Fig. 3(c)] associated with a small kink in the intensity [red circles in Fig. 3(a)]. No trace of such transition in cooling mode is observable.

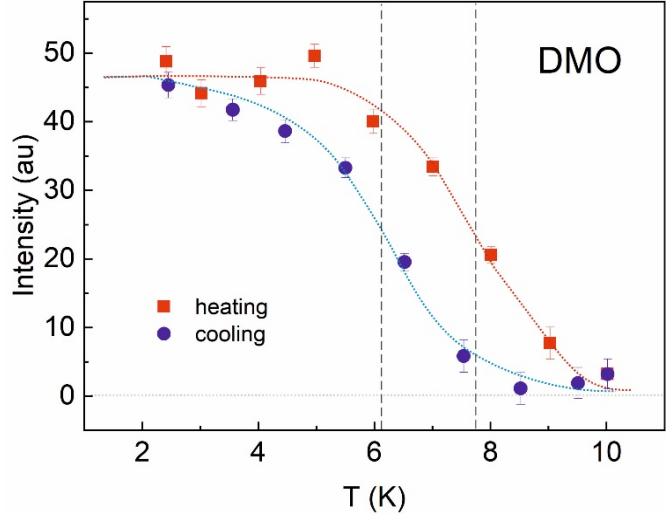

FIG. 4. Temperature dependences of integrated intensity of CM AFM reflection (0 0.5 2) in DyMnO$_3$. Solid lines are guides for the eyes.

The studies of the thermal evolution of the magnetic structure of DHMO sample were performed on the polarized neutron diffractometer POLI with the short neutron wavelength λ = 0.897 Å in non-polarized mode. The magnetic ordering in DHMO takes place at same temperature like in DMO of ~ 38 K with similar propagation vector $k^{Mn}$ = (0 $k_y$ 0) and slightly higher initial propagation vector $k_y$ = 0.367(2). In the same way, as for the DMO case, it is convenient to trace it via the temperature dependence of the satellite (2 $k_y$ 1) [Fig. 3 (d-f)]. The intensity of this magnetic satellite increases gradually below $T_N^{Mn}$ down to about 26 K, below which it remains constant until the crossover point ~ 16 K (note it coincides with $T_{CE}$ for DHMO). Below that temperature the intensity starts a much faster increase while cooling. This behavior is similar to that observed in DMO sample below 19 K due to the induced RE ordering, just happening at slightly lower temperature (ΔT ~ 3 K). The absence of the significant intensity hysteresis between the heating and cooling mode in DHMO, as well as no sharp change in the intensity of the satellite down to 4 K (lowest temperature available within POLI cryostat) [Fig. 3(d)], indicates that during our experiment in DHMO sample we always remained in the "Mn-controlled" state. Neither thermal evolution of the wave vector, no that of the FWHM in DHMO show thermal hysteresis below 16 K, demonstrating reversible thermal effect on the magnetic structure in the state where RE is mostly polarized by Mn. Above 16 K, however, the propagation vector shows a significant hysteresis between cooling and heating.

Wave vector component $k_y$ keeps constant value of 0.367(3) while cooling from 38 K down to ~ 26 K. Below this temperature, $k_y$ increases rapidly and becomes equal to 0.382(2) at $T_{Ch}$ ≈ 16 K and then it remains constant. One can suppose that this temperature $T_{Ch}$ corresponds to the transition from non-chiral spin wave type structure to cycloid, which is chiral. In the heating mode, however $k_y$ does not change its value up to ~ 26 K, then gradually decreases to the initial value $k_y$ = 0.367(3) [Fig. 3(e)]. The observed two special temperature points: 26 K and 16 K are well corresponding to the trend-change temperature points from the intensities graph [Fig. 3(d)] and those of peak width [Fig. 3(f)]. At the same time, the temperature dependence of the satellite peak-width does not show any noticeable hysteresis. Just like in the case of DMO above 19 K. Nonetheless, it is worth noting that the FWHM increases (almost doubling) between 26 K and 16 K. We relate these features to the strong interaction of the two magnetic systems – manganese and rare earth ones in the Ho substituted sample.

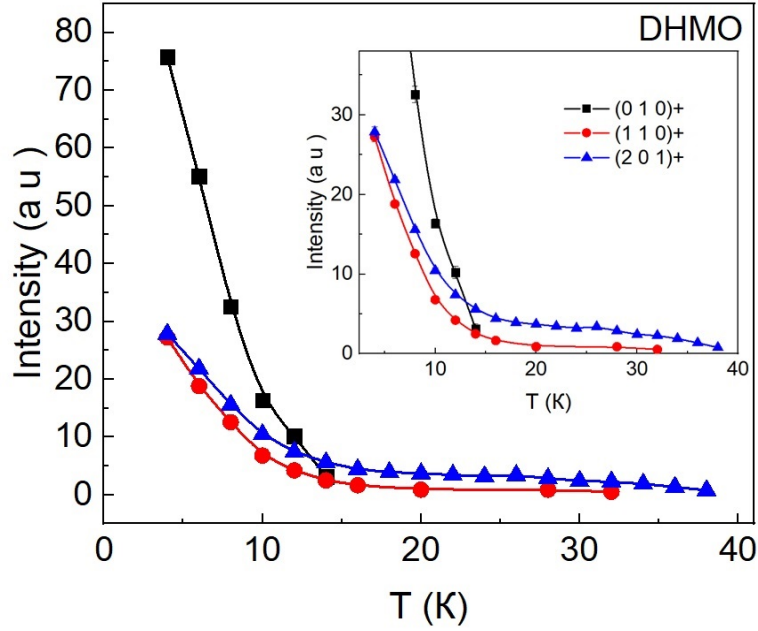

FIG. 5. Temperature evolution of the magnetic satellites of different parity (as defined in the main text) in Dy$_{0.8}$Ho$_{0.2}$MnO$_3$.

The orbit of the manganese ions in the unit cell corresponds to position 4b with sufficiently high-symmetric coordinates: Mn1 (½ 0 0), Mn2 (½ 0 ½), Mn3 (0 ½ ½), Mn4 (0 ½ 0). This gives the opportunity to obtain a correspondence between the parity type of Miller indices and magnetic Bertaut modes [39] for the manganese subsystem, as it was done in work [14]. This means, that to the reflections of type h+k – even, l – odd only A-mode, and to the reflections of type h+k – odd, l – even only C-mode, to those with h+k – even, l – even – only F-mode, and to those with h+k – odd, l – odd – only the G-mode will give a contribution. This is not quite precise for the incommensurate magnetic wave vector $\boldsymbol{k}_{Mn}$ = (0 $k_y$ 0), but nevertheless permits to make some meaningful estimations. Just below $T_N \approx$ 38 K satellites of type A emerge, originating from manganese ordering [Fig. 5], i.e. only the manganese magnetic subsystem contributes to these peaks. Similar behavior was observed for DMO in the current work as well as in previous works [14]. At $T \sim$ 32 K the satellites of type F appear, signifying the additional contribution to the magnetic ordering, which is also similar to situation at DMO [40]. Thus, it is reasonable to assume that these satellites originated from the weakly polarized rare earth subsystem. At $T \sim$ 16 K a new set of satellites corresponding to type C (h+k – odd, l – even) emerges and their intensities increases rapidly [Fig.5]. The appearance of type C reflections should be associated with the polarization between two sublattices and resulting changes in the RE sublattice.

### C. Spherical neutron polarimetry

The SNP measurements on DHMO were performed in order to distinguish between different possible magnetic structure models and to have some quantitative reference point for the estimation of the chiral scattering. Experiments with SNP technique were performed on the diffractometer POLI with neutron wavelength λ = 0.897 Å. Polarization matrix elements are determined as

$$P_{ij} = (I_{ij}^+ - I_{ij}^-)/(I_{ij}^+ + I_{ij}^-) \qquad (1)$$

where $i, j$ – directions of the initial and final polarization, $I_{ij}^+$ - intensity, measured after the scattering with final polarization along $j$, $I_{ij}^-$ - intensity, measured after the scattering with final polarization opposite to $j$. The full polarization matrices for a number of magnetic satellites of different type were measured in DHMO at 4 K. Table II shows the results for satellites (2 0.382 1) and (1 1.382 0) respectively. For the magnetic satellite of A-type (2 0.382 1), the elements of the polarization matrix $P_{yy}$ and $P_{-zz}$ are close to 1, while elements $P_{-yy}$ and $P_{zz}$ are close to -1. These elements could be expressed as [29]:

$$P_{yy} = P_{-zz} = -P_{zz} = -P_{-yy} = (M_y M_y^* - M_z M_z^*)/(M_y M_y^* + M_z M_z^*) \quad (2)$$

where $M_y, M_z$ – components of the effective magnetic interaction vector $\boldsymbol{M}_\perp = (\boldsymbol{e} \times \boldsymbol{F}_M \times \boldsymbol{e})$, with $\boldsymbol{e}$ – scattering vector, $\boldsymbol{F}_M$ – magnetic structure factor.

This unambiguously proves that the z component of the magnetic moment is very small (if any), so basically the moment is pointing along $y$ direction, which is close to the crystal axis $b$ for this particular reflection where mostly the magnetic moment of the $Mn^{3+}$ positions contributes.

The diagonal components of the polarization matrix for the F-type magnetic satellite (1 1.382 0) $P_{yy}$, $P_{-zz}$, $P_{-yy}$ and $P_{zz}$ have small absolute values (see Table II). While non-diagonal elements $P_{zy}$, $P_{-zy}$, $P_{-yz}$ are close to -1 or to 1. These elements can be expressed in the following way [29]:

$$P_{zy} = P_{yz} = -P_{-zy} = -P_{-yz} = 2Re(M_y M_z^*)/\boldsymbol{M}_\perp \boldsymbol{M}_\perp^* \quad (3)$$

Inspecting the experimental values related to the expressions (2) and (3) for the reflection (1 1.382 0) lead us to the conclusion that there should be a non-zero magnetic moment along the $z$-axis. For the scattering geometry of the satellite (1 1.382 0) the $z$-axis corresponds exactly to the crystallographic $c$-axis. Since F-type reflections originate from the RE magnetic sublattice, this statement refers to the magnetic moments of $Dy^{3+}/Ho^{3+}$ ions.

The chiral non-diagonal terms $P_{yx}$, $P_{zx}$, $P_{-yx}$ and $P_{-zx}$ of all matrices are equal to 0 within the measurement error. In this case, one can assume that the structure is either not chiral or it has an equal population of right and left-handed chiral domains. We assume that the latter is applicable in our case.

TABLE II. Measured polarization matrices for magnetic satellites (2 0.382 1) and (1 1.382 0) at 4 K in DHMO.

|  |  | (2 0.382 1) Pout | | | (1 1.382 0) Pout | | |
|---|---|---|---|---|---|---|---|
|  |  | x | y | z | x | y | z |
| Pin | x | -0.98(5) | 0.16(6) | 0.12(6) | -0.95(6) | 0.12(8) | 0.01(7) |
|  | y | -0.01(5) | 0.91(5) | -0.06(6) | 0.02(6) | 0.30(5) | -0.62(8) |
|  | z | 0.05(5) | -0.09(5) | -0.90(7) | 0.01(5) | -1.0(1) | -0.26(7) |
| Pin | -x | 0.98(5) | 0.01(5) | 0.03(5) | 0.95(6) | -0.08(7) | 0.01(7) |
|  | -y | -0.00(5) | -0.92(4) | 0.07(5) | -0.08(6) | -0.28(5) | 0.98(9) |
|  | -z | -0.04(4) | 0.05(5) | 1.04(8) | 0.02(6) | 0.81(9) | 0.31(7) |

## D. Refinement of magnetic structure from single crystal neutron diffraction data - DyMnO3

In addition to the nuclear peaks, collection of the magnetic Bragg reflections in DMO was performed at 2.4 K on diffractometer HEiDi. The data set for the commensurate ordering with wave vector $k_{Dy}$ = (0 0.5 0) consists of 250 reflections, while the one for the incommensurate order with $k_{Mn}$ = (0 0.405 0) consists of 150 reflections. Representational analysis shows, that for wave vector $k_{Dy}$ = (0 0.5 0) the RE site 4c splits into two orbits having the same basic vectors. For $Dy^{3+}$ ions we use the combination of two Γ2 representations with magnetic moments laying in the *ab* plane in the generalized configuration GxAy. It gives satisfactory convergence of the FullProf fit with $\chi^2$ = 8.4 and the resulting moment value $M_{Dy}$ = (3.51(1), 6.62(6), 0) $\mu_B$ [see Table III, Fig. 6], which perfectly matches to results (from Refs 14, 33) obtained by the neutron powder diffraction.

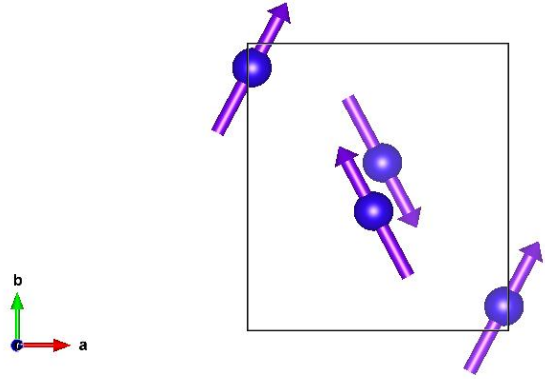

FIG. 6. Magnetic structure model of DyMnO3, only commensurate AFM order on Dy atoms at 2.4 K is shown for sake of clarity.

As for the incommensurate structure refinement at 2.4 K, for the first attempts we tried to make the refinements with Mn sublattice only, supposing that the Dy subsystem does not give any contribution to the incommensurate order. For the Mn site 4b representation Γ4b decomposes into four irreducible representations Γ1, Γ2, Γ3, Γ4. The ordering of the manganese spins was considered as a combination of two irreducible representations Γ2 and Γ3 with configuration AyAz, as it was proposed previously (in Refs. 33, 38). However, this did not lead to a convincing result in terms of fit convergence. In the next step, we included $Dy^{3+}$ sublattice in the fit as well. For $k_{Mn}$ = (0 ky 0), $k_y \neq 0.5$, representational analysis shows that site 4c splits into two orbits, similar to the case of wave vector $k_{Dy}$ = (0 0.5 0). Thus, $Dy^{3+}$ ordering with configuration GxAy has been added. For elliptical envelope this ensured a good convergence of the fit with $\chi^2$ = 2.71 and the resulting moments values: $M_{Mn}$ = (0, 2.92(26), 2.77(21)) $\mu_B$, and $M_{Dy}$ = (2.49(8), 5.57(7), 0) $\mu_B$ [Table III, Fig. 7(a)].

TABLE III. Magnetic structures of DMO and DHMO

| | Temperature | Magnetic moment | Wave vector | Mode | $M_x, \mu_B$ | $M_y, \mu_B$ | $M_z, \mu_B$ |
|---|---|---|---|---|---|---|---|
| DMO | 2.4 K | Mn | (0 $k_y$ 0) | AyAz | 0 | 2.92(26) | 2.77(21) |
| | | Dy | (0 $k_y$ 0) | GxAy | 2.49(8) | 5.57(7) | 0 |
| | | Dy (CM) | (0 0.5 0) | GxAy | 3.51(1) | 6.62(6) | 0 |
| | 12 K (colling) | Mn | (0 $k_y$ 0) | AyAz | 0 | 3.81(14) | 1.97(17) |
| | | Dy | (0 $k_y$ 0) | GxAy | 1.49(6) | 2.23(7) | 0 |
| | 12 K (heating) | Mn | (0 $k_y$ 0) | AyAz | 0 | 3.74(11) | 1.81(16) |
| | | Dy | (0 $k_y$ 0) | GxAy | 1.37(9) | 2.18(6) | 0 |
| DHMO | 4 K | Mn | (0 $k_y$ 0) | $A_yA_z$ | 0 | 3.471(15) | 0.147(13) |
| | | Dy | (0 $k_y$ 0) | GxAyAz | 1.922(15) | 4.531(15) | 0.950(11) |
| | | Ho | (0 $k_y$ 0) | GxAyAz | 4.786(15) | 7.141(14) | 4.173(12) |

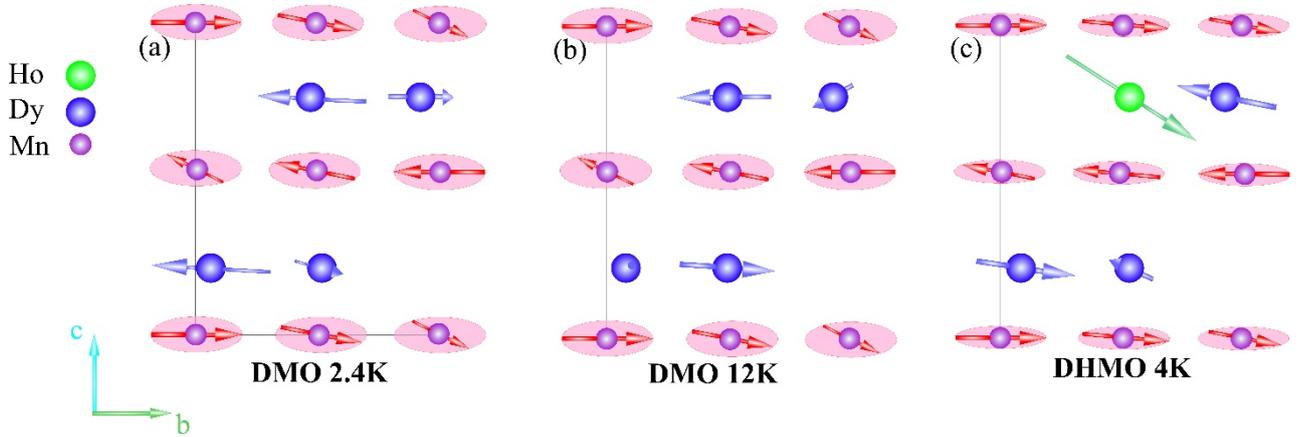

FIG. 7. Incommensurate magnetic structure model (a) DyMnO$_3$ at 2.4 K, (b) DyMnO$_3$ at 12K, (c) Dy$_{0.8}$Ho$_{0.2}$MnO$_3$ at 4 K.

For the next step of the investigation of magnetic structure evolution, we made two data collections at $T$ = 12 K, approaching this temperature in cooling and heating mode. As it can be seen in figures 3 (a-c), due to a significant difference in the satellite intensity and value of propagation vector, it is natural to anticipate different magnetic structures, realized at this temperature in two modes. Judging on the temperature dependences at Fig. 3a-c, it was naturally to suppose that for the structure obtained in heating mode there is only a little (if any) contribution from RE subsystem. For this purpose, about 200 magnetic reflections were collected for both temperature modes. We discuss first the results in cooling mode. Same as in the previous case (2.4 K) we use for refinement configuration AyAz for Mn$^{3+}$ ordering and GxAy for Dy$^{3+}$. This results in the magnetic moments values: $M_{Mn}$ = (0, 3.81(14), 1.97(17)) $\mu_B$ and $M_{Dy}$ = (1.49(6), 2.23(7), 0) $\mu_B$. Note, that obtained overall value of 2.23(7) $\mu_B$ for the Dy$^{3+}$ moment at 12 K agrees well with that of 2.5 $\mu_B$ obtained in [33]. We began the refinement of the magnetic structure at 12 K in the heating mode considering the Mn moments in AyAz configuration only, without RE contribution. However, this did not lead to a reasonable result. Instead, the inclusion of Dy magnetic system (GxAy) provided good convergence of the fit. Surprisingly, the obtained magnetic moment values in the heating mode almost do not differ from those for the cooling: $M_{Mn}$ = (0, 3.74(11), 1.81(16)) $\mu_B$, $M_{Dy}$ = (1.37(9), 2.18(6), 0) $\mu_B$. Thus, one received the same magnetic structure for both heating and cooling modes at 12 K [Fig. 7(b)]. At the same time, the temperature hysteresis [Fig. 3 (a-c)] seems to disagree to this result. We believe, that reason for this contradiction is related to the different measurement times used for data collections. In the first case, by the measurement of temperature dependence, the scan of a few satellites at one temperature-point takes just few minutes, including time for the temperature stabilization. Altogether it takes about 2 h to make the full cycle of measurements from 2.4 K to 40 K in heating mode and in particular, about 1 h to make such measurements from 2.4 K to 12 K. Contrarily to that, in the second case, for the measurement of a large number of reflections for magnetic structure refinement, full data set collection takes about 24 hrs. Thus, one can suppose that in our case of DMO we deal with a very slow magnetic structure relaxation at the fixed intermediate temperature (above $T_N^{Dy}$ and below $T_{CE}$), when the total magnetic structure during some hours transforms from its "Dy controlled" state into the "manganese controlled" one. The interesting fact is also, that at $T$ = 2.4 K Mn$^{3+}$ refined magnetic moment has a noticeably lower value than at 12 K: 2.92(26) $\mu_B$ against 3.81(14) $\mu_B$. At the same time, significant increase of satellite peak width [Fig. 3(c)] could indicate the correlation length reduction. From another point of view, the broadening of the satellites could be attributed also to the coexistence of few propagation vectors with close *k*-values. We cannot distinguish between these two possibilities, but both of them surely could be caused by the impact of the RE subsystem on the Mn one. Disturbing influence of the RE probably leads also to the reduction of the effective ordered moment found of the Mn ions at low temperatures.

## Dy$_{0.8}$Ho$_{0.2}$MnO$_3$

The data set for the refinement of the magnetic structure in DHMO was measured on diffractometer POLI with wavelength λ = 0.897 Å in non-polarized mode using lifting counter. About 150 reflections, corresponding to incommensurate propagation vector $k_{Mn}$ = (0 0.38 0) were collected at 4 K. The crystal structure of DHMO found to be the same as in DMO, therefore the representation analysis made for DMO can also be applied to DHMO. Similar as for DMO we used the model of elliptical spin cycloid for the refinement of magnetic structure in DHMO sample. However, the magnetization measurements (will follow below) and SNP results (one should pay attention on the consideration of polarization matrix elements of (1 1.38 0) satellite in Chapter C) shows the existence of a *c*-component of the magnetic moment in the RE system. Therefore, Γ2+Γ3 representation used for the manganese, has been used also for the RE subsystem. Mn magnetic order of AyAz configuration and RE order of GxAyAz give satisfactory fit with $\chi^2$ = 2.83 in contrast to $\chi^2$ = 7.11 resulting for RE configuration of GxAy only. The refined magnetic moments values are: $M_{Mn}$ = (0, 3.471(15), 0.147(13)) $\mu_B$, $M_{Dy}$ = (1.922(15), 4.531(15), 0.950(11)) $\mu_B$, $M_{Ho}$ = (4.786(15), 7.141(14), 4.173(12)) $\mu_B$ [see Table III, Fig. 7(c)]. Comparing to DMO at 12 K, *b*-component of the magnetic moment on Mn ion is only a little smaller, while *c*-component is significantly reduced in DHMO. The lower value of Mn magnetic moment can be attributed to the impact of Ho doping in RE subsystem, the disturbing influence of which increases with temperature decrease, as it can be judged by gradual increase of the satellite width [Fig.

3(f)]. Magnetic moments of the Dy ions have value a little lower than that obtained for DMO at low temperatures. The value of Ho moment in DHMO (8.27 $\mu_B$) is slightly higher than those ones reported in the literature for pure HoMnO$_3$: 7.27 $\mu_B$ [9], 7.7 $\mu_B$ [42] and is closer to free ion value of 10 $\mu_B$.

### E. Measurement of chiral scattering under applied electric fields in DHMO

Polarized neutron diffraction without polarization analysis was used in order to trace the magnetoelectric coupling in the DHMO multiferroic phase. The measurements of chiral scattering dependence from the external electric field were performed with applied voltage up to + 7.5 kV (i.e. field ~ 24.2 kV/cm) at temperatures 16 K – 20 K (close to the $T_{CE}$ in DMO). This electric field value corresponds to the maximum reachable on the sample with a positive polarity without an electric breakdown. Only -2.5 kV could be applied in the opposite polarity resulting in some field "asymmetry" in the experimental setup.

The chiral scattering was measured without polarization analysis after the sample. Initial neutron polarization was directed along the scattering vector and could be flipped – thus the magnetic scattering intensities $I_x^+$ and $I_x^-$ were measured. The difference of these intensities yield the chiral scattering:

$$I_x^+ = I_M + I_{Ch}, \quad I_x^- = I_M - I_{Ch} \qquad (4)$$

$$2I_{Ch} = I_x^+ - I_x^- \qquad (5)$$

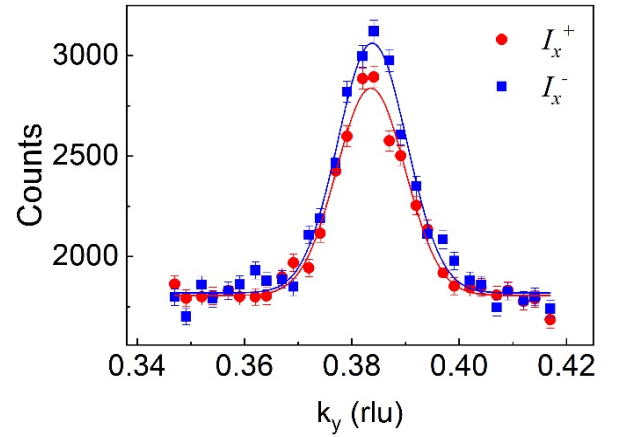

where $I_{Ch}$ - the chiral contribution to the scattering. Using this method the measurements of the magnetic chirality as function of applied electric field were performed on the satellite (2 0 1)$^+$. Small, but clearly visible difference $I_x^+ - I_x^-$ could be identified at 16 K in heating mode even with zero electric field [Fig. 8]. This observation may serve as a direct confirmation to the fact, that the magnetic structure in DHMO at 16 K is of the chiral type and some slight imbalance between the right- and left-handedness chiral domains exists in the sample at this temperature.

FIG. 8. Q-scans of magnetic satellite (2 0 1)$^+$ in DHMO sample performed with neutron polarization along scattering vector ($I_x^+$) and opposite to it ($I_x^-$) at 16 K and zero electric field.

The field measurements were performed with the external electric field applied along the crystal $c$ axis and the scattering asymmetry was calculated as

$$A = (I_x^+ - I_x^-)/(I_x^+ + I_x^-) \qquad (6)$$

representing so-called "chiral ratio", denoting the ratio of the chiral scattering to the total magnetic scattering. The measurements were made in the heating mode, each time the temperature increase was made at the maximum positive field +7.5 kV. Then, after the temperature stabilization, the measurements were performed with decreasing field down to -2 kV. After that, the field was raised again to the maximum positive value to create a close field loop and then a new temperature was established. Figure 9 displays the results of the measurements at different temperatures. It can be seen that the maximum chiral ratio value with negative sign is observed at 16 K [Fig. 9]. It does not change its overall sign when changing the field, but kind of hysteretic behavior can be interpreted to the data in dependence on the field direction change. This means that on the one side, the mobility of the chiral domains is restricted, but on the other side the magnetic chirality can be significantly changed by an external electric field at the fixed temperature depending on the field history. By the heating to 18 K the average asymmetry is dropped by a half, losing its directional hysteresis, while still remaining negative. At the higher temperatures: 19 K and 20 K it changes the sign becoming slightly positive and having rather the same values at both 19 and 20 K respectively. The field dependence of the asymmetry at these temperatures has only a week hysteretic nature (if any), which could be connected with chiral fluctuations. One can suppose that average chirality goes from negative to slightly positive because of specific way of measurements, when the measurements at each temperature were ended at the highest positive voltage, then temperature raised at this highest voltage. Thus, the measurements of the chiral scattering with polarized neutrons confirms that phase transition mentioned above in the chapter 3b at about 16 K in DHMO should be related to the formation of the chiral (cycloidal) magnetic order on Mn. As results from the refinement of the magnetic moment on the Mn in the chapter 3d this cycloid is strongly elliptic at 4 K.

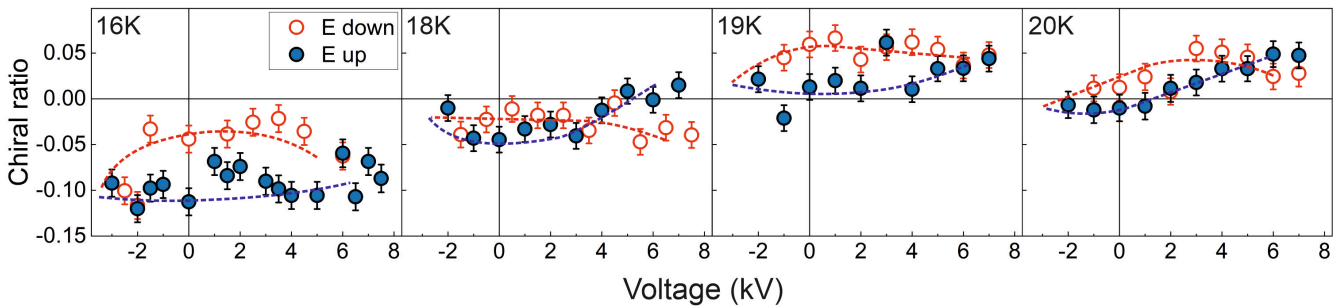

FIG. 9. The chiral scattering dependence on the external electric field in DHMO sample at temperatures between 16 K and 20 K.

The reverse of the electric field along the *c* axis favors the formation of the domain with opposite chirality (rotation direction of the spins in the cycloid) over the inverse DMI mechanism. The direct influence of the electric field on the magnetic chirality in single crystals of $TbMnO_3$ was observed by the polarized neutron scattering [11, 12] and polarized light techniques (second harmonic generation) [43]. It was possible to switch between the two magnetic chiral domain populations by cooling the sample in an external electric field over the transition temperature [11] or even by varying the external field at constant temperature. The latter however, only at temperatures sufficiently close to the ferroelectric transition temperature $T_{CE}$ [12]. By further cooling the coercive field increase significantly denoting almost invers proportional dependence on the temperature. The observed hysteresis loops of the chiral scattering dependence from electric field resembles very close the field dependence of the electric polarization by the low frequency switching. Both the polarization and chiral ratio however, do not increase significantly with temperature decrease, reaching certain saturation limit just 3 - 4 K below the $T_{CE}$. This denotes the strong pinning of the chiral domains and thus, reduced influence of the Mn-Mn DMI induced polarization at the lower temperatures. Principally similar behavior was observed also for the DMO single crystals [44]. Here however much smaller chiral ratios could be accounted and the hysteresis loop is much more flat resembling well the behavior of the electric polarization at zero magnetic field in this compound. The switching between different chiralities with moderate electric fields of few kV/cm is hardly possible (only within about 1 K close to the $T_{CE}$ of about 20 K). This shows much stronger coupling between Mn and RE magnetic sublattices in DMO comparing to the TbMnO3 and the importance of the exchange-striction mechanism in the occurring of the high electric polarization under applied magnetic field in the DMO. Our results on electric field dependence [in Fig. 9] for DHMO at 16 K are very similar to those observed for DMO at 18 K [44]. This denotes that Ho substitution at the level of about 20% on Dy position further increases the RE influence on the Mn sublattice confirming the initial hypothesis, that Ho doping controls not only the RE magnetic ordering solely, but the entire spin structure of the compound. The impossibility to switch the chirality sign in the DHMO sample at 16 K on the one side could be interpreted by the increased requested coercive field due to the Ho doping compared to the pristine DMO. On the other side also used here asymmetric field of -8.1 /+ 24.2 kV/cm may have been insufficiently high especially at the negative value. The former polarization studies on the pure DMO [45] shows the hysteresis loops of $\boldsymbol{P}_c$ with coercive fields from 26 kV/cm up to 40 kV/cm in the temperature range 18.8 K - 15.8 K respectively. In our studies, the highest electric field we could achieve in the positive polarity was of E ~ 24.2 kV/cm is close to the reported 26 kV/cm. Even if this can't be considered as an evident prove that Ho doping increases the coercive field, it can serve as a cross check that it is definitely not decreasing due to the Ho substitution.

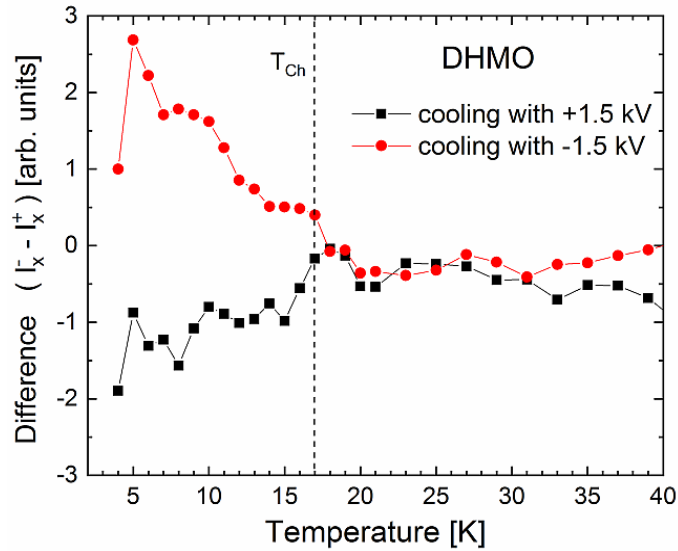

FIG. 10. The thermal dependence of the chiral scattering in DHMO sample by cooling with poling voltage of different polarity. The experimental error bars of about 2.8 arb.u. are not shown for clarity.

In order to prove whether magnetic chirality changes with temperature, sample was cooled from 45 to 4 K ones with poling voltage of +1.5 kV and ones with opposite poling voltage of -1.5 kV. $I_x^+$ and $I_x^-$ scattering intensities were measured at certain temperatures and the difference $I_x^+ - I_x^-$, denoting magnetic chirality, is built. Fig. 10 shows the measured thermal evolution. The measured chiral signal is indeed very small (at the level of the experimental standard deviation uncertainty), therefore it would be not reliable to conclude about the absolute chirality value. However, considering the data evolution clear systematics could be observed. At the temperatures above certain value $T_{Ch}$, the chiral scattering is very close to zero and seems to be independent on the poling voltage polarity. Below this value of about 17 K the chiral scattering increase with temperature decrease and have opposite sign for the two opposite polarities of the poling voltage. This observation clearly proves that reported previously for pure TbMnO$_3$ [12] and DMO [44], as well as for other multiferroics of type II [18, 46], control of magnetic chirality by external electric field is present also in Ho doped DHMO samples.

## F. Magnetization measurements

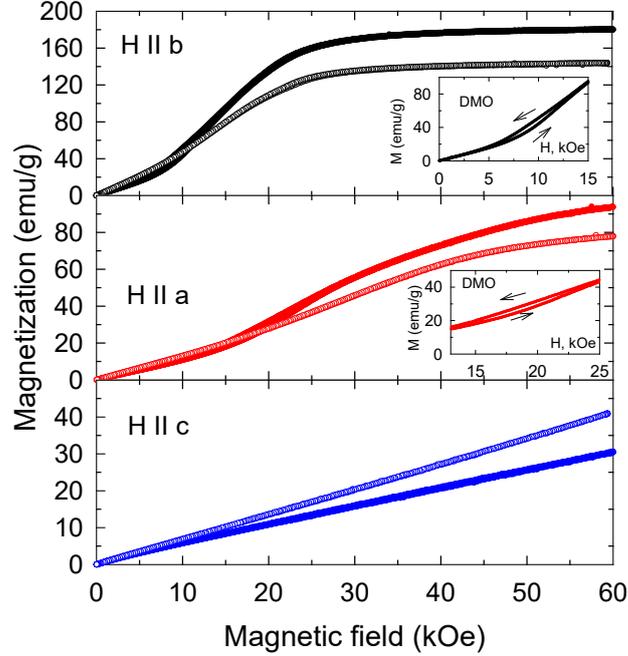

FIG. 11. Field dependences of the magnetization in single crystals of DMO (closed circles) and, DHMO (open circles) respectively measured along main crystallographic directions at T=4.2 K. Insets: hysteresis portions of DMO curves on an enlarged scale. Arrows show the variation in the external field direction.

The field dependences of the magnetization in DMO and DHMO measured along different crystallographic directions at $T = 4.2$ K are shown [in Fig. 11]. Easy to observe, that both crystals studied in this work have a significant magnetocrystalline anisotropy demonstrating quasi easy-plane behavior. However, a significant in–plane anisotropy is also present. The magnetization measured with field along $b$ axis is greater than that along $a$ axis. In DMO rare earth moments lie in the $ab$ plane and the estimation of the ratio between $a$ and $b$ components gives the direction of the easy axis, which makes ~ 29° to $b$ axis. Remarkably the ratio between $x$ and $y$ components of Dy moments at 4 K (Table III) gives almost the same angle, which is also in the good agreement with that one (from Ref. 14). It can be seen from [Fig. 11] that in DHMO value of the magnetization along the $c$ axis becomes greater than that in DMO. This is also in a good agreement with the results of magnetic structure refinement from neutron data (shown at Table III), where rare-earth magnetic moments component along $c$ axis are zero for DMO and have non-zero values for DHMO. That means that a partial replacement of Dy with Ho leads to a decrease in the magnetic moment along both the $a$ and $b$ axes and induces component along $c$.

## G. Electric polarization measurements

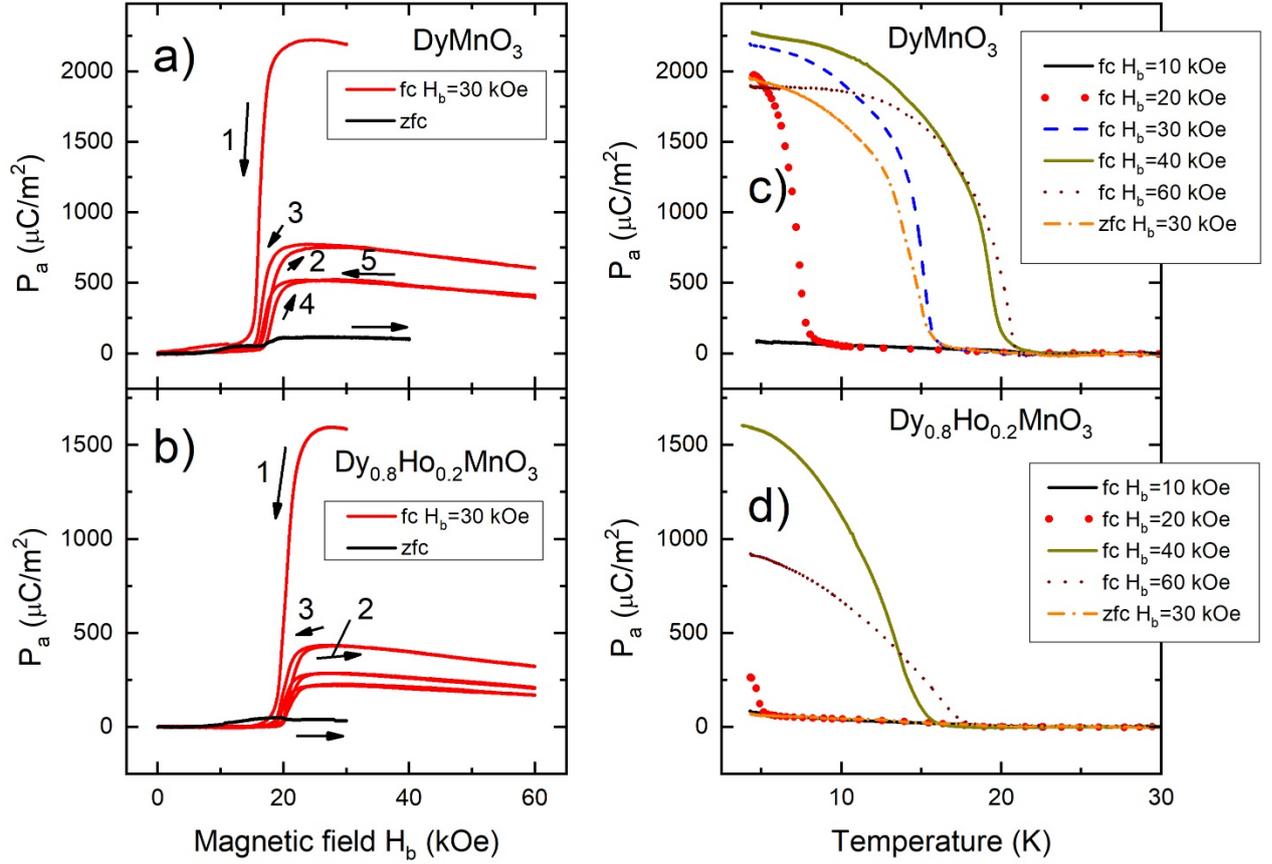

FIG. 12. Magnetic field induced electric polarization along the a axis $P_a$ in DMO [47] a) and DHMO b) measured at 4.2 K respectively. The direction of magnetic field variation during the measurements is marked by the numbered arrows, in the correspondence with the sequence of the measurements. Temperature dependences of $P_a$ at various magnetic fields for DMO [47] c) and DHMO d).

Spontaneous electric polarization in DMO of about 1400 μC/m² induced by magnetic transition to the chiral phase (at $T \sim 19$ K) is directed along the $c$ axis [1-3]. An external magnetic field applied along the $a$ or $b$ axes leads to the switch of the electric polarization vector $\boldsymbol{P}_c \rightarrow \boldsymbol{P}_a$, that can be connected with the change of magnetic structure (flop of the Mn cycloid plane from the $bc$ into $ab$ crystallographic plane) [1-3, 47]. This switch process is arising in the both magnetic field orientations $\boldsymbol{H}_b$ and $\boldsymbol{H}_a$, but in the case of $\boldsymbol{H}_b$ this occurs at the smaller field values and lead to the higher polarization value $P_a$ of more than 2000 μC/m². Here we present the results for the measurements of the polarization $P_a$ only due to some limitations in the used samples geometry and expected higher absolute polarization values. Figures 12 a and b shows the field dependences of the magnetoelectric polarization $P_a(H_b)$ for the crystals DMO [48] and DHMO respectively. The red curves were measured after pre-cooling the sample from 100 to 4.2 K in both electric $\boldsymbol{E}_a = 7.4$ kV/cm and magnetic $\boldsymbol{H}_b = 30$ kOe fields. Cooling in the magnetic field $\boldsymbol{H}_b$ leads to a reorientation of the vector of spontaneous polarization $\boldsymbol{P}_c \rightarrow \boldsymbol{P}_a$ [1-3, 48]. The electric field $\boldsymbol{E}_a$ was applied to ensure that the resulting ferroelectric state $\boldsymbol{P}_a$ is single domain one. After reaching the temperature of liquid helium, the poling voltage was switched-off and the measurement process starts. For both compounds, the polarization obtained after this prehistory reaches large values of $\boldsymbol{P}_a \sim 2200$ and $\sim 1600$ μC/m² for the pure and substituted samples, respectively. This is in a perfect agreement to the previously reported values for DMO from the single crystal measurements [3]. When the magnetic field is decreased down to $H_b \sim 17$ kOe in DMO and to $H_b \sim 20$ kOe in DHMO, a sharp drop in polarization occurs, which is associated with the re-switching of the polarization direction $\boldsymbol{P}_a \rightarrow \boldsymbol{P}_c$ (the course of the curves is indicated by the number 1 in Fig. 12 (a, b)). The following switching $\boldsymbol{P}_c \rightarrow \boldsymbol{P}_a$ occurs again by the magnetic field increasing above $H_b$~20 kOe with some coercive field hysteresis of about 1.5-2 kOe. Whereat, the coercive field strength in the substituted sample is smaller than in the DMO. This observation on the ferroelectric domains somehow correlates to the observed hysteresis loops with similar coercivity in the magnetization behavior of pure DMO [inset to Fig.11] and less pronounced magnetization hysteresis for the DHMO sample, denoting strong coupling between the magnetic and ferroelectric domains (multiferroic domains [12, 43]).

After repeated field increase, the polarization $P_a$ does not reach the previous values (curves 2). We attribute this behavior to the absence of a poling electric field $E_a$. With each subsequent iteration, the resulting $P_a$ polarization decreases, and the crystal begins to "forget" the preferred $P_a$ direction originally set by the external electric field, i.e. a domains with the different direction of the polarization tends to an equilibrium. This behavior is similar for both pristine and substituted compounds. The black curves [in Fig. 12 (a, b)] correspond to polarization measured after cooling the sample in an electric field $E_a$ = 7.4 kV/cm without applying an external magnetic field. In this case, when a magnetic field $H_b$ is applied, the sample, that initially has spontaneous $P_c$ polarization also experiences a $P_c \rightarrow P_a$ transition. Small increase of $P_a$ is observed at about 10 kOe in both samples, whereas here DHMO has slightly higher values. In the region of ~20 kOe different behavior occurs: in DMO sample $P_a$ further increase with field, while in the DHMO sample polarization lowering could be accounted. During this transition, the external electric field $E_a$ is already removed and the sample enters the multi-domain state $P_a$, in which there are domains with polarization directed along $+a$ and $-a$. This results in a compensation for the macroscopic polarization of $P_a$, the sample demonstrates overall lower polarization. Our results shows, that in the substituted sample frustrated RE magnetic ordering reacts much more flexible on the external disturbing field, than the "more rigid" well established Dy ordering in pristine DMO in good agreement with the proposed from the neutron diffraction scenario of "Dy-controlled" and "Mn-controlled" magnetic states. This assumption fit well to the observed smaller hysteresis in DHMO. Previously published results on Ho substitution in $Dy_{1-x}Ho_xMnO_3$ powders have shown overall polarization increase with increasing level of substitution between $x = 0 \div 0.3$ from about 250 to 550 $\mu C/m^2$ at low temperature (2 K) and zero applied field [23]. From the comparison of the red curves [in Fig. 12 (a, b)], it is seen however, that the polarization $P_a$ in DMO is indeed larger comparing to the one in substituted DHMO. It is worth noting that it is hardly possible to directly compare the overall spontaneous polarization in zero field from the powder sample with the field-induced polarization along a certain direction in the single crystal. As shown in Ref. [23] the application of the magnetic field of 30 kOe reduces the accounted polarization in the x = 0.1 DHMO powder sample almost double from about 400 to 200 $\mu C/m^2$. In our case, such a field strength is just necessary to create much larger $P_a$ in the single crystal. So, a direct comparison between our results from the pristine and substituted single crystals do not confirms the previously reported absolute polarization value enhancement due to the Ho substitution.

The noticeable differences in the behavior of ferroelectric polarization of DMO and DHMO are observed at the temperature dependences [Fig. 12 (c) [48] and (d), respectively]. Measurements of temperature dependencies were made in the heating mode after cooling with the simultaneously applied electric $E_a$ and magnetic $H_b$ fields, as described above, except for the zfc curves [shown in orange in Fig. 12 (c, d)], which will be discussed later. As shown in Fig. 12 (c and d), after cooling in the $H_b$=10 kOe, the polarization $P_a$ increase only slightly reaching similarly large values for both samples. This field is not sufficient to induce the polarization flop of $P_c \rightarrow P_a$. In the 20 kOe field-cooled samples the $P_a$ reaches significant values in DMO, transition happening at about 7 K. In DHMO however, only the beginning of the $P_c \rightarrow P_a$ transition is induced by $H_b$ = 20 kOe at $T$ = 4.2 K [Fig. 12 (d)]. The shift of 3-4 K toward lower temperatures is observed. As the temperature increases, the polarization undergo reverse transition $\boldsymbol{P}_a \rightarrow \boldsymbol{P}_c$ after which the curve of polarization dependence coincides with the curve measured in $\boldsymbol{H}_b = 10$ kOe. A further increase of the magnetic field leads to a shift of the $\boldsymbol{P}_a \rightarrow \boldsymbol{P}_c$ transition towards higher temperatures up to the transition temperature $T_{Ch} \sim 20$ K for DMO and about 16 K for DHMO respectively, still maintaining the observed low temperature shift of about 4 K. This value is in a good agreement to that determined from the thermal evolution of the magnetic structure by the neutron diffraction and polarization analysis. For the DMO, the ferroelectric state is more stable and continues to exist in larger magnetic fields up to temperatures even exceeding temperature of the magnetic transition $T_{Ch}$ defined in the absence of a magnetic field. One should also mention the curves shown in orange in Fig. 13 c and d. They were measured after cooling in zero magnetic and electric fields down to a temperature of 4.2 K, after which a constant poling field $\boldsymbol{E}_a$ and an external magnetic field $\boldsymbol{H}_b = 30$ kOe were applied to the samples. After that, the electric field was removed, and measurements of temperature dependences of $\boldsymbol{P}_a$ were made while heating. As [shown in the Fig. 12 (c)] after this prehistory, the polarization in the DMO is quite large and almost reaches the values obtained after cooling in the field. A completely different picture is observed for the substituted sample DHMO. For that crystal, the polarization obtained after such a prehistory does not differ from the polarization in the 10 kOe field, which indicates that the $\boldsymbol{P}_c \rightarrow \boldsymbol{P}_a$ transition is not induced. This means that also no flop of the Mn cycloidal magnetic order plane is produced in the DHMO, denoting that Ho substitution not only lead to some frustration of magnetic order on the RE position, but also directly influencing the Mn magnetic sublattice as well, e.g. by strengthening the RE-Mn interactions.

## IV SUMMARY AND CONCLUSIONS

DMO is improper multiferroic of the type II and shows one of the highest electric polarization for these type of compounds. This behavior is caused by the presence of two emergent mechanisms for electric polarization: the inverse DMI on the chiral Mn sublattice and the exchange striction between the RE and Mn sublattices. The latter is stronger than the former and it depends strongly on the temperature. This is expressed in the fact that with a decrease in temperature and the appearance of an intrinsic independent long-range order of Dy subsystem, the RE-RE coupling is enhanced while RE-Mn one is weakened and the electric polarization also significantly decreases. It was shown previously in powder samples [23, 49] that doping of Dy by Ho at a level of 0.2 – 0.3 suppresses this effect and leads to an increase in the electric polarization. The observation was attributed to the increased, by doping, disorder of the RE subsystem, maintaining thus the RE-Mn interaction to lower temperatures. In order to understand how this process works at microscopic level and also to test the proposed mechanisms of multiferoicity in this phase, we carried out a comprehensive comparative study of magnetic ordering in single crystals of unsubstituted DMO and substituted DHMO with a doping level of 0.2 using neutron diffraction methods of both unpolarized and polarized neutron beams. Preliminarily, using neutron diffraction on single crystals, the crystal structure of the samples was determined at room temperature (DHMO) and at low temperatures (DMO, DHMO). To study the role of electromagnetic coupling at the microscopic level and its relationship with the chirality of the magnetic structure on substituted samples, experiments were carried out by the method of polarized neutron diffraction with the application of external electric fields at different temperatures. To determine the relationship between the experimentally determined microscopic parameters of the magnetic structure and their influence on the behavior of electric polarization in these compounds, an additional measurement of the electric polarization in external magnetic fields was carried out at the same temperatures and values of electric fields as for microscopic studies. Macroscopic and microscopic studies were carried out on identical samples obtained from the same crystal grow badge to exclude the sample dependence effect. The obtained data for DMO are compared in detail with previously published in the literature, as well as to the data on DHMO single crystals, which are presented for the first time.

Our results on the low-temperature crystal structure in DMO shows that reported for the room temperature crystal structure symmetry within orthorhombic space group *Pbnm* describes well the average structure also at very low temperatures in different magnetic phases. The refined at 2.4 K structural parameters like lattice parameters, atomic coordinates and both isotropic and anisotropic thermal displacements are presented in supplemental materials. The structural parameters of the partially substituted $Dy_{0.8}Ho_{0.2}MnO_3$ both at room and low temperatures (4 K) are reported for the first time. The results show that the overall crystal symmetry of the parent compound with *Pbnm* space group is maintained also in the substituted samples. Ho occupies the same Wyckoff positions as Dy and the lattice parameters were obtained to be very similar for both compounds. No significant magnetic-order-related symmetry change was observed at low temperature within accuracy of the measurements. In the same time doping by 20% of Ho significantly changes the magnetic structure, introducing additional disorder in the RE sublattice. This is reflected in the fact that AFM ordering temperature of RE in DHMO is shifted towards lower values comparing to DMO. Detailed investigation of the temperature-evolution of the magnetic structures in DMO and DHMO both in cooling and heating modes confirms the existence of strong magneto-crystalline coupling but also magnetic RE-Mn coupling in the both compounds. This coupling, more pronounced in DMO, leads to the occurrence of two temperature-dependent magnetic states: "Mn-controlled" and "Dy-controlled" and pronounced temperature hysteresis. Different temperature hysteresis is observed in DHMO, indicating weaker influence of the RE magnetic subsystem on Mn in the substituted DHMO. The precise values of the ordered magnetic moments on Dy and Mn positions respectively were obtained from the magnetic structure refinement in both phases: above and below $T_N^{Dy}$ in DMO. Our results clearly demonstrate that strong RE-Mn coupling continue to persist also below $T_N^{Dy}$. An attempt to observe a hysteretic dependence in the microscopic magnetic order depending on the heating or cooling history at 12 K showed no difference in the magnetic ordering. We attribute this to the possible presence of the very slow magnetic relaxation dynamics due to the strong RE-Mn coupling. It is known from previous investigations [3] that Mn forms chiral strongly elliptical cycloid structure in *bc* plane below 19 K in DMO. Our results reproduces well this physical picture, however we show that after ordering of Dy in CM AFM structure below $T_N^{Dy}$ of about 7 K the elipticity of the Mn structure reduces significantly, creating almost circular cycloid. This may serve as trace for the reduction of the inter-coupling between the Mn and Dy sublattices in DMO below $T_N^{Dy}$. In the Ho doped HDMO sample however, strongly elliptical envelope of the Mn order (almost collinear along crystal *b* direction) is observed at 4 K. The chiral character of the magnetic order of Mn in DHMO is very weak. But using polarized neutron diffraction its appearance below 17 K could be unambiguously shown. SNP measurements and complete polarization matrices for incommensurate magnetic structure in DHMO are reported for the first time. These measurements provide an important hint for the magnetic structure model in DHMO, as they evidence the presence of RE-magnetic moment component along *c* direction, which is not the case for the undoped sample. Thus, doping by 20% Ho on Dy position transforms RE magnetic subsystem from "more ordered" 2D-components character to the "less ordered" 3D-components type. That leads to situation when two magnetic subsystems, manganese and rare earth ones have a coherent incommensurate spatial propagation. Resulting magnetic structure is favorable for the generating ferroelectric polarization originating from both inverse DMI at Mn subsystem and RE-Mn exchange striction mechanisms in whole temperature region of cycloidal ordering.

Interestingly, not only the moment on the minority Ho dopants are pointing differently in regard to those of the majority Dy matrix, but also Dy became differently ordered than in "pure" DMO. Our polarization measurements on the doped single crystals do not demonstrate reported for powder samples increase in the polarization value due to the Ho substitution. At this point we would like to mention that it is hardly possible to make direct comparison of the overall spontaneous polarization from the powder with the field-induced polarization along a unfovarable direction in the single crystal. By the application of electric field along $c$ axis, we could change the population of the chiral domains in DHMO to a some small extent at constant fixed temperature in the vicinity of $T_{Ch}$. This indicates a rather strong pinning of the chiral domains to the crystal frame. Our results on electric field dependence for DHMO at 16 K are very similar to those observed for DMO at 18 K [44]. This denotes that Ho substitution at the level of about 20% on Dy position further increase the RE influence on the Mn sublattice confirming the initial hypothesis, that Ho doping controls not only the RE magnetic ordering solely, but the entire spin structure of the compound.

## ACKNOWLEDGMENTS

The authors are grateful to S.V. Gavrilov for the technical assistance. The authors are grateful to D.A. Balaev for fruitful discussions and crystal growth management. This work was supported by the Russian Foundation for Basic Research grant # 19-52-12047, and DFG grant # SA 3688/1-1, also by Russian Foundation for Basic Research, Government of Krasnoyarsk Territory, Krasnoyarsk Regional Fund of Science to the research projects number 18-42-243024 and 20-42-243008. The neutron data were obtained at the instruments HEiDi and POLI operated jointly by RWTH Aachen University and Jülich Centre for Neutron Science within JARA cooperation.

---